\documentstyle[12pt]{article}
\title{ O(a) perturbative  improvement for Wilson fermions}

\author{Satchidananda Naik\thanks{Alexander-von-Humboldt Fellow}
\\
Max-Planck-Institut f\"ur Physik \\
Werner-Heisenberg-Institut \\
F\"ohringer Ring 6 \\
D-8000 Munich 40}

\begin{document}
\maketitle

\newcommand{\bee}{\begin{equation}}
\newcommand{\nn}{\nonumber}
\newcommand{\ee}{\end{equation}}
\newcommand{\ba}{\begin{array}}
\newcommand{\ea}{\end{array}}
\newcommand{\bea}{\begin{eqnarray}}
\newcommand{\eea}{\end{eqnarray}}
\newcommand{\e}{\epsilon_+(i\kappa)}
\newcommand{\eps}{\epsilon}
\newcommand{\pa}{\partial}
\newcommand{\lb}{\lbrack}
\newcommand{\Se}{S_{\rm eff}}
\newcommand{\rb}{\rbrack}
\newcommand{\de}{\delta}
\newcommand{\th}{\theta}
\newcommand{\ka}{\kappa}
\newcommand{\al}{\alpha}
\newcommand{\bt}{\beta}
\newcommand{\si}{\sigma}
\newcommand{\bsi}{\Sigma}
\newcommand{\vp}{\varphi}
\newcommand{\g}{\gamma}
\newcommand{\gb}{\Gamma}
\newcommand{\om}{\omega}
\newcommand{\kh}{\hat k}
\newcommand{\ph}{\hat p}
\newcommand{\omb}{\Omega}
\newcommand{\pr}{\prime}
\newcommand{\ra}{\rightarrow}
\newcommand{\nb}{\overline N}
\newcommand{\MSb}{{\overline {\rm MS}}}
\newcommand{\lnh}{\ln(h^2/\Lambda^2)}
\newcommand{\df}{\delta f(h)}
\newcommand{\h}{{1\over2}}
\newcommand{\R}{m/\Lambda}
\newcommand{\abschnitt}[1]{\par \noindent {\large {\bf {#1}}} \par}
\newcommand{\subabschnitt}[1]{\par \noindent
                                          {\normalsize {\it {#1}}} \par}
\newcommand{\skipp}[1]{\mbox{\hspace{#1 ex}}}

%
%
%
%
\newcommand\dsl{\,\raise.15ex\hbox{/}\mkern-13.5mu D}
\newcommand\delsl{\raise.15ex\hbox{/}\kern-.57em\partial}
\newcommand\Ksl{\hbox{/\kern-.6000em\rm K}}
\newcommand\Asl{\hbox{/\kern-.6500em \rm A}}
\newcommand\Dsl{\hbox{/\kern-.6000em\rm D}} 
\newcommand\Qsl{\hbox{/\kern-.6000em\rm Q}}
\newcommand\gradsl{\hbox{/\kern-.6500em$\nabla$}}
%
\begin{abstract} \normalsize

The coefficients of the $O(a)$ -improved Sheikholeslami-Wohlert
action for Wilson
fermions are perturbatively determined at one-loop level
and estimated at two-loop level.

\end{abstract}

\newpage
\pagestyle{plain}
\setcounter{page}{1}
\subabschnitt{1. Introduction}

   Several groups are investing considerable effort to numerically
calculate various hadronic decay constants and the mass spectrum
with Wilson fermions \cite{UKQ}.
One of the systematic errors, that due to the
finite lattice spacing, can in principle be diminished
 by taking an
improved action \`{a} la Symanzik \cite{SYMLW}.
In the weak coupling expansion the pure Yang-Mills theory has no
O(a) cutoff effects in the infinite volume or
with a  finite   volume with periodic boundary conditions,
however the fermionic part of the Wilson action introduces such
terms. To eliminate these
 effects all independent  dimension five operators
are added to the action. The procedure drastically simplifies
by demanding that  only the  on-shell physical quantities
 of the theory are improved
\cite{LW}.
Then only the
operator $\bar{\psi}(x){\sigma}_{\mu\nu}P_{\mu\nu}(x)\psi(x)$
needs to be added to the action
where $P_{\mu\nu}$ is the term with four plaquettes touching
at point $x$  in the form of  a `clover leaf'
, an equivalent of $F_{\mu\nu}$ \cite{SW}.
The fermionic part of the action reads
\bea
&& S^W_F~= a^4 {\sum}_{x,\mu}~\lb \bar{\psi}(x)\lb (r-{\g}_{\mu}) U_{\mu}(x)
\psi (x+\mu a) + (r + {\g}_\mu ) {U_{\mu}}^{\dag}(x-\mu a) \psi(x-\mu a)
  \rb \nn \\
&&~~~~~~~~
+{i\over 2} c({g_0}^2){\sum}_{\nu}
\bar{\psi}(x){\sigma}_{\mu\nu}P_{\mu\nu}(x)\psi(x)\rb \nn
\\
\eea
with
\bee
P_{\mu \nu}(x) ={1\over 8} \lb U^P_{\mu ,\nu}(x) + U^P_{-\mu ,\nu}
 (x) +U^P_{\mu,-\nu}(x)+U^P_{-\mu ,-\nu}(x)~ -~ (\mu \ra \nu) \rb
\ee
where
$U^P_{\mu,\nu}(x)$ is the untraced plaquette variable.
The coefficient $c(g_0^2)$ ought to be fixed perturbatively to
all orders of $g^2_0$.
This coefficient happens to be
quite important for the hadronic decay amplitude and for the hadron
spectrum when the short distance effects play a role e.g.
for the hyperfine splitting in charmonium.
At tree level, the improvement condition fixes this to be `r'
the Wilson parameter \cite{SW}.
The one-loop coefficient was first calculated by Wohlert\footnote{
The action given in Ref.\cite{W} is not exactly the clover action
now used in numerical simulations. We found terms missing
from the fermion-three gluon vertex function and errors
in the expression of Clebsch-Gordon coefficients. This created
a suspicion of a drastic change of the result since it involves
the most dominant tadpole graph. We made an independent check however
and found that the number quoted is correct and these errors
are probably just misprints.}
 \cite{W} and also there is a mean field estimate
 given by Lepage and Mackenzie \cite{LM}. In this article
 we independently check this coefficient at the one loop level
and give an  approximate value in the two loop level.

\subabschnitt{2. General Strategy}
To fix the coefficient $c(g^2_0)$ to all orders of perturbation
theory a suitable on-shell quantity needs to be defined.
The most obvious of such quantities which instantly comes to mind is
 to look for the pole of the fermion two point function.
 This however unfortunately cannot be related to $c(g^2_0)$
 due to  the   Slavnov-Taylor identity
\bee
-i{{\pa}\over{\pa p_{\mu }}} \bsi (p) = {\gb }_{\mu} (p, p, 0) ,
\ee
and  the fact that the improved part of  the  vertex function
$ {\gb }_{\mu }(p,p,k)~ (\propto {\sum}_{\nu}
{}~{\si}_{\mu\nu}~sin k_\nu)
 $ vanishes for $k~=~0$.
 This clearly shows that to all orders of
perturbation  $c(g^2_0)$ is not sensitive to the on-shell
fermion two point function. Another quantity of interest
is $<\bar{\psi}(y)P e^{ig_0{\int}^x_y A_\mu (\xi)d{\xi}^{\mu}}\psi(x)>$,
however does not yield a  tree level improvement condition.
 We were so far unable to find a suitable
 quantity which needs an
improvement  at tree level in the infinite volume. However,
one possibility is to assume that our lattice
has finite extent L in $x_1$ and $x_2$ directions with twisted
periodic boundary conditions.
 Due to these
 boundary conditions, quarks and gluons get mass by a Kaluza-Klein
type of mechanism and some of the eigenvalue modes of the
transfer matrix
   remain stable for small
coupling in this twisted world. These states are created from the
vacuum by gauge invariant operators such as a  Wilson loop
winding around the torus.
The spectrum of the
S-matrix for these gauge invariant
  modes in the LSZ scheme can be studied
unambiguously
 as has been already done for the
pure Yang-Mills theory to fix two necessary coefficients \cite{LW}.
A  more  detailed discussion of
this procedure is given in Ref.\cite{LW}.

The twisted periodic  boundary condition for the gauge field
reads,
\bee
U(x+L_{\tilde{\nu}}, \mu) = \omb_{\nu}U(x,\mu){\omb_{\nu}}^{-1},~
\nu = 1, 2,
\ee
where $\omb_{\nu}$ are constant SU(N) matrices with
\bee
\omb_1 \omb_2 = e^{{2\pi i}\over N} \omb_2 \omb_1 .
\ee
To use the twisted boundary condition for  fermions
one introduces an extra internal degree of freedom ``smell''
\cite{PS} and a fermion  belongs to the ${\bar{N_S}} \times
N_C$ representation. The twisted antiperiodic  boundary condition
on the torus reads
\bee
\psi (x+L_{\nu}) = \omb_{\nu} \psi(x) {\omb_{\nu}}^{-1}
e^{{i\pi}\over N}~~\nu =1,2.
\ee
Thus the transverse momenta
in these directions take discrete values as
\bee
p_{\perp} =( 2 n_1 +1, 2 n_2 +1)m,~~~~~ n_i \in Z ,
\ee
where  $m = {\pi \over{LN}}$. This offers a mass gap even though
we have a zero bare mass fermion to start with. The Fourier decomposition
of the gauge field reads
\bee
A_{\mu}(x) = {1\over{L^2N}} {\sum}_{k_{\perp}} {\int}_{k_0,k_3}
\exp (ikx) ~{\gb}_k ~\exp {{ik_\mu}\over 2}{\tilde{A}}_{\mu}(k)
\ee
where $k_{\perp} = 2m n_{\nu}$ and $\gb_k$ plays the role of
SU(N) group generator. Since $A_{\mu}(x)$ is traceless, ~
  there exists no zero mode.
The gluon propagator reads
\bee
D_{\mu \nu}(k)= -\h ~{\chi}_k~ Z(k,k)\lb {\de}_{\mu \nu} \kh^2
+(\al -1) \kh_\mu  \kh_\nu \rb {1\over{(\kh^2)^2}},
\ee
where
\bea
  {\chi }_k =& 0 ~ if ~ k_{\perp } =0, ({\rm mod N} ) \nn \\
            =&1 ~~{\rm otherwise} ,\nn\\
\eea
$\al$ is the gauge fixing parameter
 and $\kh_\mu~=~ 2~ sin {{k_\mu}\over 2}$. Also
\bee
Z(k, k') = z^{\h(< k, k'> -(k , k'))}
\ee
where $z = e^{{2\pi i}\over N} $ ,$<k , k'> =n_1 {n_1}'+n_2 {n_2}'
+(n_1+n_2)({n_1}'+{n_2}')$ and $(k,k')~ = ~ n_1{n_2}'-n_2 {n_1}'$.
Following Ref.\cite{LW} we
 study the spectrum of the LSZ scattering  process (c.f. Fig.1).
The   on-shell momenta for these fermions
are $p_1 = (iE, m,m,im)$ , ${p_1}'=(iE,m,-m,-im)$ ,
 $p_2=(iE,m,-m,-im)$ and ${p_2}' = (iE,m,m,im)$.
This kinematical
choice of momenta
simplifies the calculation drastically since the exchange diagram
Fig.1b does not contribute and also the fermion wave function
renormalization  to $O(a)$  does not contribute to the S-matrix elements.
Following Ref. \cite{LW},
we look for the residue of the pole of the scattering amplitude
\bee
S=T_{\mu} Z^2_G(k) D_{\mu \nu}(k) {T_{\nu}}',
\ee
where $T_{\mu}~=~ {\bar U }(p_1)~ {\gb}_{\mu}(p_1,{p_1}',k)~U({p_1}')$
and $Z_G(k)$ is the gluon wavefunction renormalization.
Also ${T_{\nu}}'$  can be obtained by replacing $p_1\ra p_2$ , ${p_1}'
\ra {p_2}'$  and $k~\ra -k$ in $T_{\mu}$.  The residue of the pole of this
S-matrix element;
\bee
\lambda = z(-k,{p_1}') T_{\mu}{T_\mu}'.
\ee
We assume here a perturbative expansion of  this residue
$\lambda~=~{\sum}_{i=1}^{\infty}~ {g_0}^{2i}~ {\lambda}^{i} $
and also for
$T_\mu~=~{\sum}_{i=0}^{\infty}~ {g_0}^{2i+1}~{T_\mu}^{i} $
and  $c(g_0^2)~=~c_0~ +g_0^2 c_1~ +g_0^4 c_2~+~\ldots $.

To illustrate the case in the tree level
\bee
{(T^0_\mu)}_{\al \bt} = T^0_{\mu}(1) + T^0_{\mu}(a) \\
=z(k,p) {\bar U}_{\al}(p)\lb -{\g}_\mu - {i\over 2}(c_0 - r)(p +p')_\mu
 + O(a^2) \rb U_{\bt}(p') .
\ee
We look for the residue in  a particular polarization of the gluon
say $\mu ~=~1$ and in the fixed helicity state of the fermion
say $\al ~= \bt =~1$. For the zero bare mass fermion $U(p)~=~
(\sinh E )^{-1} .({\g}_0 \sinh E~+~ i {\g}_i {\ph}_i)~U^1 $,
where $U^1_\bt~=~{\de}_{1,\bt}$.
This gives
\bea
&& T^0_1(1)~=~2iz,\quad ~~~~~T^0_1(a)~=~-2izm(c_0-r)\nn \\
&&{T^0_1(1)}'(1)~=~-2i,\quad{T^0_1(a)}'~=~2im(c_0-r)\nn \\
\eea
 and
\bea
&&{\lambda}^0(a)~=~(T^0_1(1)~{T^0_1}'(a)+~T^0_1(a){T^0_1}'(1)) \nn \\
&&~~~~~~~~=~-8zm(c_0-r).\nn \\
\eea
To demand the tree level improvement we set ${\lambda}^0 (a)~=~0 $
and  thus
get the condition $c_0~=~r$ \cite{SW}.

To all orders of perturbation theory   $T_1(a)$ gets contributions
from the fermion wave function renormalization, vertex function
renormalization and also a term coming from the naive expansion
of $c(g^2_0)$. So
\bee
  T_1(a)~=~ T_1^c~+T_1^{WF}~+T_1^{VF}.
\ee
 However for our kinematical choice of momenta,
it can be proved that
\bee
T^0_1(1)~{T^{WF}_1}'(a)~+ T^{WF}_1(a)~{T^0_1}'(1)~=0.
\ee
The contribution of the
gluon wave function renormalization $Z_G(k)$ does not contribute
to  $\lambda $  for the $O(a)$ improvement.
So it remains to calculate only $T_1^{VF}(a)$ and ${T^{VF}_1}'(a)$
to all orders of perturbation. The three point function
${\bar U}(p)~ {\gb}_{\mu}(p,p',k)~U(p')$  is expressed as
${\bar U}^1~ {\sum}_i~ O_i~ B_i~U^1$ where $O_i$'s are sixteen
 bilinear invariant  Dirac    basis
 matrices   and $B_i$ 's are their coefficients.
It is sufficient to look for the residue in  the particular channel
which we have chosen, here $\propto {\si}_{12}$ for every diagram.

\subabschnitt{3. $O(g^2)$ Improvement}
To one loop order there are six diagrams (c.f.Fig.2) contributing
to $T_1(a)$.
Due to the twisted periodic boundary conditions  all these
loop integrals depend on L and N. Then the contributions are compared
with their asymptotic expansion
\bee
I \approx {\sum}^{\infty}_{i=0}({\al}_i~+{\bt}_i~ \ln~ m)(m)^{i+d}
\ee
where the lattice spacing is set to unity, $m~=~{{\pi}\over{LN}}$
and $d$ is the degree of divergence of the graph.
The method of evaluation of these graphs is exactly the
same as in Ref.\cite{LW}. The analytical expressions for the vertex
functions and loop integrands are quite lengthy and  beyond
the scope of this paper . So we present here
only the results.

The imaginary part and the coefficient of the $\log~m$ for
daigram 2.b exactly cancel with that of 2.c. This is also true for
the diagram 2.d where this part
  cancels with the sum of the contributions coming from
2.e and 2.f.\footnote{
 This can be analytically checked in the following way. We take a very small
generic external momenta $\epsilon$ for these diagrams
and then make a Taylor's expansion  in
$\epsilon$ of these loops to get the same tensor structure of all these
loops by using symmetry of the integration of the internal momenta.
(In the Feynman gauge expressions are considerably simpler.)
 Then
we look
for the logarithmic divergence term for each loop.}
 The small contributions coming from  diagram 2.b-f are not checked by us,
however we use here the results of    Wohlert \cite{W}. The contribution of
each diagram is
given in Table I. for $N=2$ and $N=3$ in the Feynman gauge. Needless
to state that the total
 contributions of these graphs to the residue of the
S-matrix is independent of the choice of the gauge.

{\bf{Table 1.    \hspace{3pt}
Values of the one-loop graphs for $N=2$ and $N=3$}}.
\vspace{10pt}
\[
\begin{tabular}{||c|c|c|} \hline\hline
                        Fig.             &
   N=2                        &
   N=3 \\ \hline
   2.a  & 0.34800  &  0.594789 \\ \hline
   2.b  & -0.01630  &  -0.03268 \\ \hline
   2.c  & 0.01512   &  0.03087 \\ \hline
   2.d  & 0.03864   &  0.05787 \\ \hline
   2.e  & -0.03903  &  - 0.05953 \\ \hline
   2.f  & -0.03903   & - 0.05953 \\ \hline
   Total & 0.308  &   0.53179 \\ \hline
\end{tabular}
\]
Here also one observes that the contribution of 2.b and 2.c
nearly cancel with each other so  also
of 2.d  nearly  cancel with 2.e .The only leftovers are the contributions of
2.a which is the most dominant one and the contribution of
  2.f is nearly 10$\% $
of 2.a. This clearly shows the evidence of the tadpole dominance
\cite{LM}.

So the residue
\bee
 {\lambda}^1(a)~=~4m~\lb 0.308 ~-~2c_1\rb ,
\ee
 for $N=2$  and
\bee
 {\lambda}^1(a)~=~4m~\lb 0.5318~-2c_1 \rb
\ee
for $N=3$.
This gives $c_1~=~0.154 $ for $N=2 $ and $c_1~=~.2659 $ for $N=3 $.

\subabschnitt{4. $O(g^4)$ Improvement}
The numerical simulations for the hadron spectrum or decay amplitudes
 are performed for $\bt~\approx ~6.$ i.e. $g^2_0~\approx~ 1.$
 So it is worthwhile also to fix the coefficient $c_2$ .
 From our observation in the previous section and also the earlier
investigation of perturbative computation of lattice graphs \cite{HW}
 there is an indication that the main contribution
comes from the tadpole graph. In our case at one loop level
 all  graphs nearly cancel
with each other except the tadpole one
 and also all are less than 10$\%$ of the latter.
 Thus  it seems legitimate to
 take only the perturbative correction from tadpoles as a first step
 towards  a full  and much lengthier
     calculation of the two loop improvement.
 The dominant
two loop graphs are given in Fig.3. and the
 results are presented in Table II.
Here again one also observes the tadpole dominance. The main contribution
comes from Fig.3.a and  the gauge invariant part of Fig.3c.
 The contribution of the latter which is
  like plaquette-plaquette  correlation function  $D_{\mu \nu,\mu\nu}$,
gives 0.1489, the largest among all these graphs.
All other contributions are less than 10$\%$ of these two quantities.
\newpage
{\bf Table II.    \hspace{3pt}
Values of the two-loop graphs for $N=2$ and $N=3$}.
\vspace{10pt}
\[
\begin{tabular}{||c|c|c|} \hline\hline
                        Fig.             &
   N=2                        &
   N=3 \\ \hline
   3.a  & 0.03403  &  0.10465  \\ \hline
   3.b  & -0.0259 &   0.00648 \\ \hline
   3.(c+d+e+f+g)  & 0.05346   &  0.16174  \\ \hline
   3.(h+i)  & $-n_f$ 0.01  &  $-n_f$  0.02  \\ \hline
\end{tabular}
\]

This gives $c_2=~0.0207$ for $N~=~2$ and $c_2=~0.1164$ for $N~= 3$
and $n_f~=~2$.
\subabschnitt{5. Conclusion}
A meanfield type of estimate
of this coefficient is given by\cite{LM}
\bee
 c(g^2_0)~=~ ({1\over N}<tr U_P>)^{-{3\over 4}}.
\ee
The perturbative expansion of this plaquette expectation value
reads \cite{GR}
\bee
{1\over N}<tr U_P> = 1 - u_1 g^2_0 - u_2 g^4_0 +\ldots
\ee
where
\bee
u_1 = {(N^2 -1)\over{8N}} , \quad u_2 = {{(N^2 -1)}\over{48 N}}\{
{{2N^2 - 3}\over{8N}} + N K_1\}
\ee
with $K_1~=~-0.0042$.
This gives $ c_1~=~.25 $ and $c_2~=~.098$ whereas we get $c_1~=~.266$
and $c_2~=~.1164$ with two flavors.
It is quite gratifying to observe that
 by two  different
methods of calculation one gets a fair agreement of the estimate of
these  coefficients, which can now be used in numerical simulations
with more confidence.

{\it Acknowledgements}

I deeply express my gratitude to P. Weisz for motivating me to do this
calculation, for numerous discusions at every stage of the work
and providing me some of his unpublished notes. I am very much
thankful to Prof. W. Zimmermann for hospitality at the Max-Planck Institute
to complete this work.
 I  also acknowledge the  Alexander-von-Humboldt
Foundation for the financial support.

\newpage
\abschnitt{Figure captions}
Fig.1. The on-shell fermion fermion scattering

Fig.2. The one-loop graphs

Fig.3. The most dominant two-loop graphs

\end{document}